\documentclass[onecolumn,3p]{elsarticle}

\usepackage{graphicx}
\usepackage{amssymb}
\usepackage{amsmath}
\usepackage{hyperref}
\usepackage[usenames]{color}

\journal{Computer Physics Communications}

\begin{document}

\begin{frontmatter}

\title{OpenMP Fortran and C programs for solving the time-dependent Gross-Pitaevskii equation in an anisotropic trap}

\author[ift]{Luis E. Young-S.\corref{author}}
\ead{luisevery@gmail.com}

\author[scl]{Du\v{s}an Vudragovi\'{c}}
\ead{dusan.vudragovic@ipb.ac.rs}

\author[bdu]{Paulsamy Muruganandam}
\ead{anand@cnld.bdu.ac.in}

\author[ift]{Sadhan K. Adhikari}
\ead{adhikari@ift.unesp.br}

\author[scl]{Antun Bala\v{z}}
\ead{antun.balaz@ipb.ac.rs}

\cortext[author] {Corresponding author.}
\address[ift]{Instituto de F\'{\i}sica Te\'{o}rica, UNESP -- Universidade Estadual Paulista, 01.140-70 S\~{a}o Paulo, S\~{a}o Paulo, Brazil}
\address[scl]{Scientific Computing Laboratory, Institute of Physics Belgrade, University of Belgrade, Pregrevica 118, 11080 Belgrade, Serbia}
\address[bdu]{School of Physics, Bharathidasan University, Palkalaiperur Campus, Tiruchirappalli -- 620024, Tamil Nadu, India}

\begin{abstract}
We present new version of previously published Fortran and C programs for solving the Gross-Pitaevskii equation for a Bose-Einstein condensate
with contact interaction in one, two and three spatial dimensions in imaginary and real time, yielding both stationary and non-stationary solutions.
To reduce the execution time on multicore processors, new versions of parallelized programs are developed using
Open Multi-Processing (OpenMP) interface.
The input in the previous versions of programs was the mathematical quantity nonlinearity for dimensionless form of Gross-Pitaevskii equation,
whereas in the present programs the inputs are quantities of experimental interest, such as, 
number of atoms, scattering length, oscillator length for the trap, etc.
New output files for some integrated one- and two-dimensional densities of experimental interest are given.
We also present speedup test results for the new programs. 
\end{abstract}


\end{frontmatter}

\begin{small}
\noindent
{\bf New version program summary}

\noindent
{\em Program title:} BEC-GP-OMP package, consisting of: (i) imag1d, (ii) imag2d, (iii) imag3d, (iv) imagaxi, (v) imagcir, (vi) imagsph, (vii) real1d, (viii) real2d, (ix) real3d, (x) realaxi, (xi) realcir, (xii) realsph. \\
{\em Catalogue identifier:} AEDU\_v4\_0.\\
{\em Program Summary URL:} \href{http://cpc.cs.qub.ac.uk/summaries/AEDU_v4_0.html}{http://cpc.cs.qub.ac.uk/summaries/AEDU\_v4\_0.html}\\
{\em Program obtainable from:} CPC Program Library, Queen's University of Belfast, N. Ireland.\\
{\em Licensing provisions:} Apache License 2.0\\
{\em No. of lines in distributed program, including test data, etc.:} 130308. \\
{\em No. of bytes in distributed program, including test data, etc.:} 929062.\\
{\em Distribution format:} tar.gz.\\
{\em Programming language:} OpenMP C; OpenMP Fortran.\\
{\em Computer:} Any multi-core personal computer or workstation. \\
{\em Operating system:} Linux and Windows.\\
{\em RAM:} 1 GB.\\
{\em Number of processors used:} All available CPU cores on the executing computer. \\
{\em Classification:} 2.9, 4.3, 4.12.\\
{\em Catalogue identifier of previous version:} AEDU\_v1\_0, AEDU\_v2\_0.\\
{\em Journal reference of previous version:} Comput. Phys. Commun. \textbf{180} (2009) 1888; {\it ibid.} \textbf{183} (2012) 2021.\\
{\em Does the new version supersede the previous version?:} No. It does supersedes versions AEDU\_v1\_0 and AEDU\_v2\_0, but not AEDU\_v3\_0, which is MPI-parallelized version.\\

\noindent\\
{\em Nature of problem:}
The present OpenMP Fortran and C programs solve the time-dependent nonlinear partial differential Gross-Pitaevskii (GP) equation for a Bose-Einstein 
condensate in one (1D), two (2D), and three (3D) spatial dimensions in a harmonic trap with six different symmetries: axial- and radial-symmetry in 3D, circular-symmetry 
in 2D, and fully anisotropic in 2D and 3D.

\noindent\\
{\em Solution method:}
The time-dependent GP equation is solved by the split-step Crank-Nicolson method by discretizing in space and time.
The discretized equation is then solved by propagation, in either imaginary or real time, over small time steps.
The method yields the solution of stationary and/or non-stationary problems.

\noindent\\
{\em Reasons for the new version:}
Previously published Fortran and C programs \cite{bec2009,bec2012} for solving the GP equation 
are recently enjoying frequent usage \cite{uca} and application to a more complex scenario of dipolar atoms \cite{dbec2015}.
They are also further extended to make use of general purpose graphics processing units (GPGPU) with Nvidia CUDA \cite{dbec2016},
as well as computer clusters using Message Passing Interface (MPI) \cite{bec2016}.
However, a vast majority of users use single-computer programs, with which the solution of a realistic dynamical 1D problem,
not to mention the more complicated 2D and 3D problems, could be time consuming. 
Now practically all computers have multicore processors, ranging from 2 up to 18 and more CPU cores.
Some computers include motherboards with more than one physical CPU, further increasing the possible number of available CPU cores on a single computer to several tens.
The present programs are parallelized using OpenMP over all the CPU cores and can significantly reduce the execution time.
Furthermore, in the old version of the programs \cite{bec2009,bec2012} the inputs were based on the mathematical quantity nonlinearity for the dimensionless form of the GP equation.
The inputs for the present versions of programs are given in terms of phenomenological variables of experimental interest,
as in Refs.~\cite{dbec2015,dbec2016}, i.e., number of atoms, scattering length, harmonic oscillator length of the confining trap, etc.
Also, the output files are given names which make identification of their contents easier, as in Refs.~\cite{dbec2015,dbec2016}.
In addition, new output files for integrated densities of experimental interest are provided, and all programs were 
thoroughly revised to eliminate redundancies.

\noindent\\
{\em Summary of revisions:}
Previous Fortran \cite{bec2009} and C \cite{bec2012} programs for the solution of time-dependent GP equation in 1D, 2D, and 3D
with different trap symmetries have been modified to achieve two goals. First, they 
are parallelized using OpenMP interface to reduce the execution time in multicore processors.
Previous C programs \cite{bec2012} had OpenMP-parallelized versions of 2D and 3D programs, together with the serial versions, while here all programs are
OpenMP-parallelized. 
Secondly, the programs now have input and output files with quantities of phenomenological interest. 
There are six trap symmetries and both in C and in Fortran there are twelve programs, six for imaginary-time propagation and six for real-time propagation, totaling to 24 programs. 
In 3D, we consider full radial symmetry, axial symmetry and full anisotropy. In 2D, we consider circular symmetry and full anisotropy.
The structure of all programs is similar.

For the Fortran programs the input data (number of atoms, scattering length, harmonic oscillator trap length, trap anisotropy, etc.) are conveniently placed at the beginning of 
each program. For the C programs the input data are placed in separate input files, examples of which can be found in a directory named input.
The examples of output files for both Fortran and C programs are placed in the corresponding directories called output.
The programs then calculate the dimensionless nonlinearities actually used in the calculation. 
The provided programs use physical input parameters that give identical nonlinearity values as the
previously published programs \cite{bec2009,bec2012}, so that the output files of the old and new programs can be directly compared.
The output files are conveniently named so that their contents can be easily identified, following Refs.~\cite{dbec2015,dbec2016}.
For example, file named $<$code$>$-out.txt, where $<$code$>$ is a name of the individual program, is the general output file
containing input data, time and space steps, nonlinearity, energy and chemical potential, and was named fort.7 in the old Fortran version.
The file $<$code$>$-den.txt is the output file with the condensate density, which had the names fort.3 and fort.4 in the old Fortran version for imaginary- and
real-time propagation, respectively.
Other density outputs, such as the initial density, are commented out to have a simpler set of output files.
The users can re-introduce those by taking out the comment symbols, if needed.

Also, some new output files are introduced in this version of programs. 
The files $<$code$>$-rms.txt are the output files with values of root-mean-square (rms) sizes in the multi-variable cases.
There are new files with integrated densities, such as imag2d-den1d\_x.txt, where the first part (imag2d) denotes that the density was calculated 
with the 2D program imag2d, and the second part (den1d\_x) stands for the 1D density in the $x$-direction, obtained after integrating out the 2D density $|\phi(x,y)|^2$
in the $x-y$ plane over $y$-coordinate, 
\begin{equation}
n_{1D}(x)= \int _{-\infty}^{\infty} dy |\phi(x,y)|^2\, .
\end{equation}
Similarly, imag3d-den1d\_x.txt and real3d-den1d\_x.txt represent 1D densities from a 3D calculation obtained after integrating out the 3D density $|\phi(x,y,z)|^2$
over $y$- and $z$-coordinate. 
The files imag3d-den2d\_xy.txt and real3d-den2d\_xy.txt are the integrated 2D densities in the $x-y$ plane from a 
3D calculation obtained after integrating out the 3D density over the $z$-coordinate, and similarly for other output files.
Again, caclulation and saving of these integrated densities is commented out in the programs, and can be activated by the user, if needed.

\begin{table}[tp]
\caption{Wall-clock execution times (in seconds) for runs with 1 and 20 CPU cores with different programs using the 
Intel Fortran ifort (F-1 and F-20, respectively) and Intel C icc (C-1 and C-20, respectively) compilers,
and obtained speedups (speedup-F=F-1/F-20, speedup-C=C-1/C-20).}
\label{tab1}
\centering
\begin{tabular}{ccccccc}
\hline
& F-1 & F-20 & speedup-F & C-1&C-20 & speedup-C \\ \hline
imag1d&32 & 27 &1.2 & 45 & 27&1.7\\
imagcir& 15 & 15 & 1.0& 21 & 15 &1.4\\
imagsph&14 & 14& 1.0 & 19 & 10 & 1.9\\
real1d& {327} & 122&{2.7} & 451 & 143 &3.1\\
realcir& {132} & 57&{2.3} & 182& 64 &2.8\\
realsph& 119 & 67&1.8 & 191 & 61 & 3.1 \\ \hline
imag2d&{190} &{  52}& { 3.7} & {394} & {33} &11.9\\
imagaxi& {240}&{  56}&{ 4.3} & {499} & {55}& { 9.1}\\
real2d& {269} &{ 47}&{ 5.7} & 483 & 35 &13.8\\
realaxi&{132} &{ 25}& { 5.3} & 237 & 22 & 10.8 \\ \hline
imag3d&{1682} & 366&4.6 & 2490& 202 &12.3\\
real3d&{15479} & 2082&7.4 & 22228 & 1438 & 15.5\\ \hline
\end{tabular}
\end{table}

\begin{figure}[!t] 
\begin{center}
\includegraphics[width=8cm]{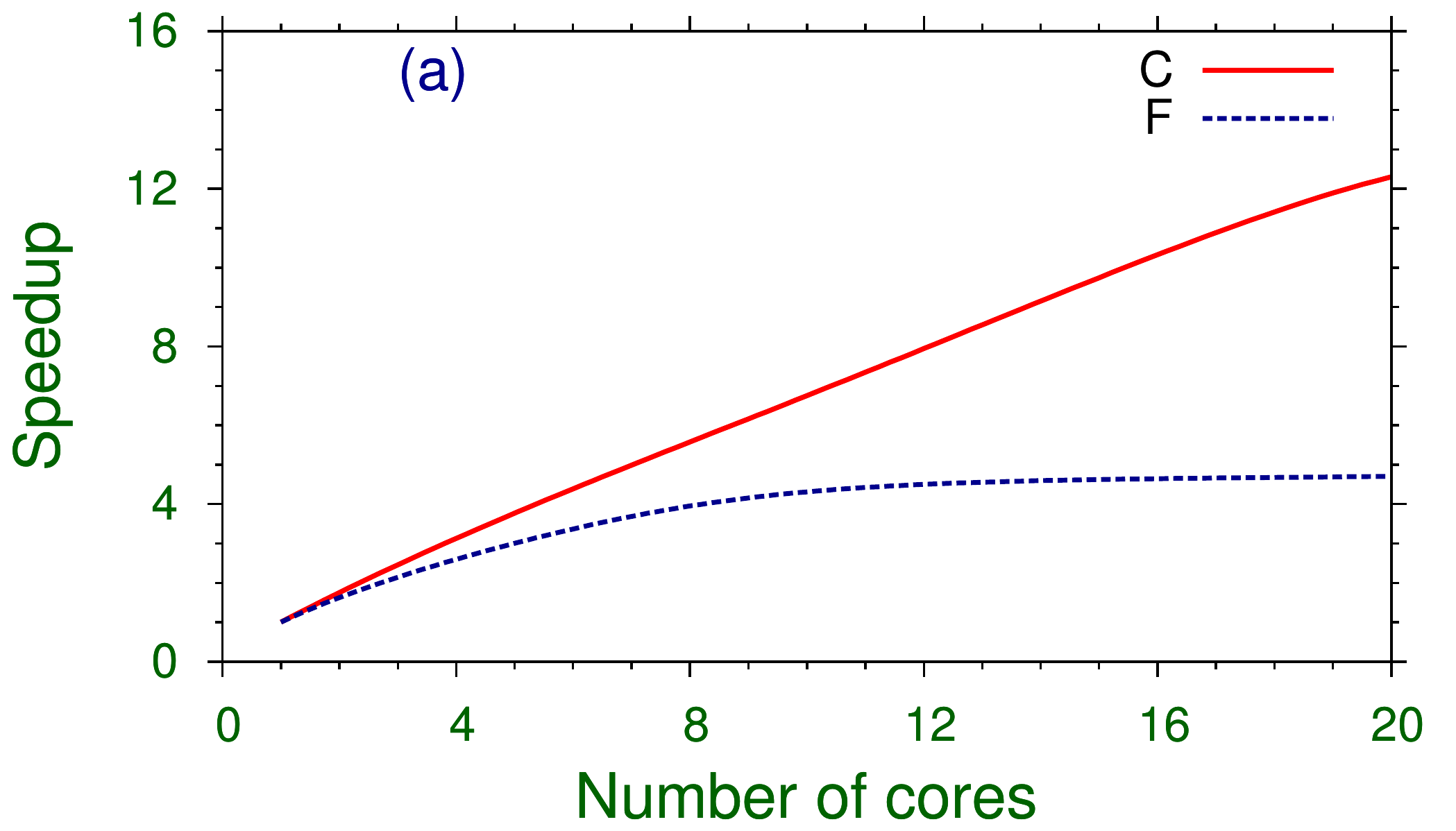}
\includegraphics[width=8cm]{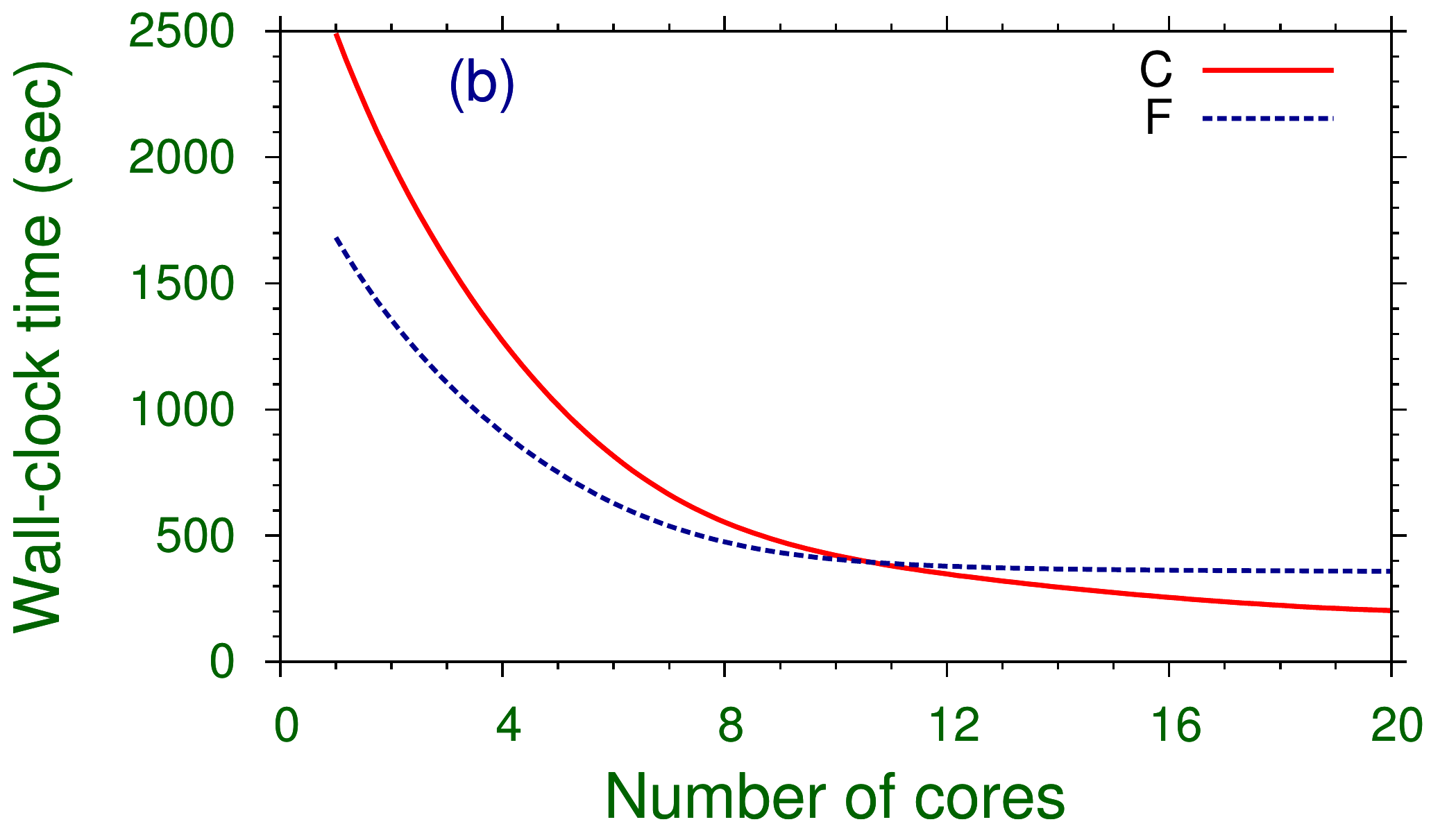}
\caption{(a) Speedup of the C and Fortran (F) imag3d programs as a function of the number of cores, measured in a
workstation with two Intel Xeon E5-2650 v3 CPUs, with a total of 20 cores.
The speedup for the run with $N$ cores was calculated as the ratio between wall-clock execution times with one and $N$ cores. (b) Wall-clock time of the same runs as a function of the number of cores.}
\label{fig1}
\end{center}
\end{figure} 
\vspace*{8mm}

In real-time propagation programs there are additional results for the dynamics saved in files, such as
real2d-dyna.txt, where the first column denotes time, the second, third and fourth columns display rms 
sizes for the $x$-, $y$-, and $r$-coordinate, respectively. The dynamics is generated by multiplying the nonlinearity 
with a pre-defined factor during the NRUN iterations, and starting with the wave function calculated during the NPAS iterations. 
Such files were named fort.8 in the old Fortran versions of programs.
There are similar files in the 3D real-time programs as well.

Often it is needed to get a precise stationary state solution by imaginary-time propagation and then use it
in the study of dynamics using real-time propagation.
For that purpose, if the integer number NSTP is set to zero in real-time propagation, the density obtained in 
the imaginary-time simulation is used as initial wave function for real-time propagation, as in Refs.~\cite{dbec2015,dbec2016}.
In addition, at the end of output files $<$code$>$-out.txt, we have introduced two new outputs, 
wall-clock execution time and CPU time for each run.

We tested our programs on a workstation with two 10-core Intel Xeon E5-2650 v3 CPUs, and present results for all programs compiled with the Intel compiler.
In Table~\ref{tab1} we show different wall-clock execution times for runs on 1 and 20 CPU cores for Fortran and C.
The corresponding columns ``speedup-F" and ``speedup-C" give the ratio of wall-clock execution times of runs on 1 and 20 CPU cores, and denotes 
the actual measured speedup.
For the programs with effectively one spatial variable, the Fortran programs turn out to be quicker for small number of cores, whereas for larger number of cores 
and for the programs with  three spatial variables the C programs are faster.
We also studied the speedup of the programs as a function of the number of available cores.
The results for the imag3d Fortran and C programs are illustrated in Figs.~\ref{fig1}(a) and \ref{fig1}(b), where we plot the speedup and actual wall-clock time of the imag3d C and Fortran programs as a function of number of cores in a workstation
with two Intel Xeon E5-2650 v3 CPUs, with a total of 20 cores. 
The plot in Fig.~\ref{fig1}(a) shows that the C program parallelizes more efficiently than the Fortran program.
However, as the wall-clock time in Fortran for a single core is less than that in C, the wall-clock times in both cases are comparable, viz.~Fig. ~\ref{fig1}(b).
A saturation of the speedup 
with the increase of the number of cores is expected in all cases. However, the saturation is attained quicker in Fortran than in C programs, and therefore the use
of C programs could be  recommended for larger number of CPU cores. For a small number of cores the Fortran programs should be preferable. 
In Fig.~\ref{fig1}(a) the saturation of the speedup of the Fortran program is achieved for  approximately 10 cores, when the wall-clock time of the C program crosses that of the Fortran program.  
 
\noindent\\
{\em Additional comments:}\\
This package consists of 24 programs, see Program title above. For the particular purpose of each program, please see descriptions below.

\noindent\\
{\em Running time:}\\
Example inputs provided with the programs take less than 30 minutes in a workstation with two Intel Xeon Processor E5-2650 v3, 2 QPI links, 10 Cores (25~MB cache, 2.3~GHz).

\noindent\\
Program summary (i), (v), (vi), (vii), (xi), (xii)\\

\noindent
{\em Program title:} imag1d, imagcir, imagsph, real1d, realcir, realsph.\\
{\em Title of electronic files in C:} (imag1d.c and imag1d.h), (imagcir.c and imagcir.h), (imagsph.c and imagsph.h), (real1d.c and real1d.h), (realcir.c and realcir.h), (realsph.c and realsph.h).\\
{\em Title of electronic files in Fortran 90:} imag1d.f90, imagcir.f90, imagsph.f90, real1d.f90, realcir.f90, realsph.f90.\\
{\em Maximum RAM memory:} 1~GB for the supplied programs.\\
{\em Programming language used:} OpenMP C and Fortran 90.\\
{\em Typical running time:} Minutes on a modern four-core PC.\\
{\em Nature of physical problem:} These programs are designed to solve the time-dependent nonlinear partial differential GP equation in one spatial variable. \\
{\em Method of solution:} The time-dependent GP equation is solved by the split-step Crank-Nicolson method by discretizing in space and time. The discretized equation is then solved by propagation in imaginary time over small time steps. The method yields the solution of stationary problems.\\

\noindent\\
Program summary (ii), (iv), (viii), (x)\\
\\
\noindent
{\em Program title:} imag2d, imagaxi, real2d, realaxi.\\
{\em Title of electronic files in C:} (imag2d.c and imag2d.h), (imagaxi.c and imagaxi.h), (real2d.c and real2d.h), (realaxi.c and realaxi.h). \\
{\em Title of electronic files in Fortran 90:} imag2d.f90, imagaxi.f90, real2d.f90, realaxi.f90. \\
{\em Maximum RAM memory:} 1~GB for the supplied programs.\\
{\em Programming language used:} OpenMP C and Fortran 90.\\
{\em Typical running time:} Hour on a modern four-core PC.\\
{\em Nature of physical problem:} These programs are designed to solve the time-dependent nonlinear partial differential GP equation in two spatial variables. \\
{\em Method of solution:} The time-dependent GP equation is solved by the split-step Crank-Nicolson method by discretizing in space and time. The discretized equation is then solved by propagation in imaginary time over small time steps. The method yields the solution of stationary problems.\\

\noindent Program summary (iii), (ix)\\
\\
\noindent
{\em Program title:} imag3d, real3d.\\
{\em Title of electronic files in C:} (imag3d.c and imag3d.h), (real3d.c and real3d.h). \\
{\em Title of electronic files in Fortran 90:} imag3d.f90, real3d.f90.\\
{\em Maximum RAM memory:} 1~GB for the supplied programs.\\
{\em Programming language used:} OpenMP C and Fortran 90.\\
{\em Typical running time:} Few hours on a modern four-core PC.\\
{\em Nature of physical problem:} These programs are designed to solve the time-dependent nonlinear partial differential GP equation in three spatial variables. \\
{\em Method of solution:} The time-dependent GP equation is solved by the split-step Crank-Nicolson method by discretizing in space and time. The discretized equation is then solved by propagation in imaginary time over small time steps. The method yields the solution of stationary problems.\\

\section*{Acknowledgements}
\noindent
L.~E. Y.-S. acknowledges support by the FAPESP of Brazil under project
2012/21871-7 and 2014/16363-8. D.~V. and A.~B. acknowledge support by the 
Ministry of Education, Science, and Technological Development of the Republic of Serbia under 
projects OI1611005, ON171017 and III43007.
P.~M. acknowledges support by the Science and Engineering Research Board, Department of Science and Technology, Government of India under project No.~EMR/2014/000644.
S.~K.~A. acknowledges support by the CNPq of Brazil under project 303280/2014-0, and by the FAPESP of Brazil under project 2012/00451-0.
 
\end{small}

\end{document}